\documentclass[aps,prl,floatfix,twocolumn,showpacs]{revtex4}
\usepackage{graphicx} \usepackage{amsmath} \usepackage{amssymb} 
\newcommand{\comment}[1]{}
\newcommand{\BEQ}{\begin{equation}}
\newcommand{\EEQ}{\end{equation}}
\newcommand{\BEA}{\begin{eqnarray}}
\newcommand{\EEA}{\end{eqnarray}}
\renewcommand{\d}{{\rm d}}

\newcommand{\zmu}{\mu}

\newcommand{\w}{{\bf w}}
\newcommand{\f}{\theta}
\newcommand{\chichi}{{\boldsymbol \chi}}
\newcommand{\etaeta}{{\boldsymbol \eta}}
\newcommand{\ff}{{\boldsymbol \theta}}
\newcommand{\nunu}{{\boldsymbol \nu}}

\newcommand{\bt}{\tilde{t}}
\newcommand{\bz}{\tilde{z}}
\newcommand{\by}{\tilde{y}}
\newcommand{\br}{\tilde{r}}

\begin{document}
\draft
\title{Explaining Zipf's Law via Mental Lexicon}
\author{Armen E. Allahverdyan$^{1,3)}$, Weibing Deng$^{1,2,4)}$ and Q. A. Wang$^{1,2)}$}

\address{$^{1)}$Laboratoire de Physique Statistique et Syst\`emes Complexes, ISMANS, 44 ave. Bartholdi, 72000 Le Mans, France\\
$^{2)}$ IMMM, UMR CNRS 6283, Universit\'e du Maine, 72085 Le Mans, France\\
$^{3)}$Yerevan Physics Institute, Alikhanian Brothers Street 2, Yerevan 375036, Armenia \\
$^{4)}$Complexity Science Center and Institute of Particle Physics, Hua-Zhong Normal University, Wuhan 430079, China
}
 
\begin{abstract} The Zipf's law is the major regularity of statistical
linguistics that served as a prototype for rank-frequency relations and
scaling laws in natural sciences. Here we show that the Zipf's law|together
with its applicability for a single text and its generalizations to high
and low frequencies including hapax legomena|can be derived from
assuming that the words are drawn into the text with random
probabilities. Their apriori density relates, via the Bayesian
statistics, to general features of the mental lexicon of the author who
produced the text. 

\end{abstract}




  
  

\maketitle

The Zipf's law states that in a given text the ordered and normalized
frequencies $f_1>f_2> ...$ for the occurence of the word with rank $r$
behave as $f_r\propto r^{-\gamma}$ with $\gamma\approx 1$
\cite{zipfwiki,baa}. This law applies to texts written in many natural
and artificial languages. Its almost universal validity fascinated
generations of scholars, but its message is still not well understood:
is it just a consequence of simple statistical regularities
\cite{mandelbrot,li}, or it reflects a deeper structure of the text
\cite{shreider_sharov}? 
Many approaches were proposed for deriving the Zipf's law suggesting
that it can have different origins.  They are divided into two
groups. 

{\it (1)} Certain theories deduce the law from certain general premises of
the language \cite{sole,prokopenko,mandelbrot,dunaev,mitra,manin}.  The
general problem of derivations from this group is that explaining the
Zipf's law for the language (and verifying it for a frequency
dictionary) does not yet mean to explain the law for a concrete text,
where the frequency of the same word varies widely from one text to
another and is far from its value in a frequency dictionary
\cite{arapov}. 

{\it (2)} The law can be derived from certain probabilistic models
\cite{shrejder,arapov,simon,zane,kanter,hill,li}.  Albeits some of these
models assume relevance for realistic text-generating processes
\cite{zane,kanter}, their a priori assumed probability structure is
intricate, hence the question ``why the Zipf's law?" translates into
``why a specific probabilistic model?" By far most known probabilistic
model is a random text, where words are generated through random
combinations of letters and the space symbol seemingly reproducing the
$f_r\propto r^{-1}$ shape of the law \cite{mandelbrot,li}. But the
reproduction is elusive, since the model leads to a huge redundancy|many
words have the same frequency and length|absent in normal texts
\cite{cancho}. 

Our approach for deriving the Zipf's law also uses a
probability model. It differs from previous models in several respects.
First, it explains the law for a single text together with its limits of
validity, i.e. together with the range of ranks where it holds.  It also
explains the rank-frequency relation for very rare words (hapax
legomena) and relates it to the Zipf's law. Second, the a priori
structure of our model relates to the mental lexicon \cite{mental} of
the author who produced the text.  Third, the model is not {\it ad hoc}:
it is based on the latent semantic analysis that is used successfully
for text modeling.

\textit{\textbf{The validity range of the Zipf's law}}. Below we
present empirical results examplified on 3 English texts [see Table~I]
that clarify the validity range of the law,
confirm known results, but also make new points
that motivate the theoretical model worked out in the sequel.

\begin{table}
\begin{center}
\tabcolsep0.05in \arrayrulewidth0.3pt
\renewcommand{\arraystretch}{0.4}
\caption{
Parameters of 3 texts: {\it The Age of Reason} (AR) by T. Paine, 1794 (the major source of British deism).
{\it Thoughts on the Funding System and its Effects} (TF) by P. Ravenstone, 1824 (economics). 
{\it Dream Lover} (DL) by J. MacIntyre, 1987 (romance novella).
Total number of words $N$, number of different words $n$,
the lower $r_{\rm min}$ and the upper $r_{\rm max}$ ranks of the Zipfian
domain, the fitted values of $c$ and $\gamma$.}
\begin{tabular}{cccccccc}
\hline\hline
Texts & $N$ &  $n$ &$r_{\rm min} $& $r_{\rm max}$ & $c$ & $\gamma$    \\
\hline
TF & 26624 &  2067 &  36  & 371& 0.168& 1.032 \\
AR & 22641 &  1706 &  32  & 339& 0.178& 1.038  \\
DL & 24990 &  1748 &  34  & 230& 0.192& 1.039   \\
\hline
\end{tabular}
\end{center}
\end{table}

For each text we extract the ordered frequencies of $n$ different words:
\BEA
\label{00}
\{f_r\}_{r=1}^{n}, ~~ f_1\geq ...\geq f_{n}, ~~ {\sum}_{r=1}^{n} f_r =1.
\EEA
To fit $\{f_r\}_{r=1}^{n}$ to the Zipf's form $\hat{f}_r=cr^{-\gamma}$,
we represent the data as $\{y_r(x_r)\}_{r=1}^{n}$, where $y_r=\ln f_r$
and $x_r=\ln r$, and fit it to the linear form $\{\hat{y}_r=\ln c-\gamma
x_r\}_{r=1}^{n}$. Two unknowns $\ln c$ and $\gamma$ are obtained from
minimizing the sum of squared errors $SS_{\rm err}={\sum}_{r=1}^{n}
(y_r-\hat{y}_r)^2$ \cite{sup}.  Now ${\rm min}_{c,\gamma}[SS_{\rm err}]=SS_{\rm
err}^*$ and the correlation coefficient $R^2$ between
$\{y_r\}_{r=1}^{n}$ and $\{\hat{y}_r\}_{r=1}^{n}$ \cite{lili,sup} measure
the fitting quality: $SS_{\rm err}^*\to 0$ and $R^2\to 1$ mean good
fitting.  We minimize $SS_{\rm err}$ over $c$ and $\gamma$ for $r_{\rm
min}\leq r\leq r_{\rm max}$ and find the maximal value of $r_{\rm
max}-r_{\rm min}$ for which $SS_{\rm err}^*$ and $1-R^2$ are smaller
than, respectively, $0.05$ and $0.005$. This value of $r_{\rm
max}-r_{\rm min}$ also determines the final fitted values of $c$ and
$\gamma$; see Table I and \cite{sup}.

\begin{figure}
\includegraphics[width=6cm]{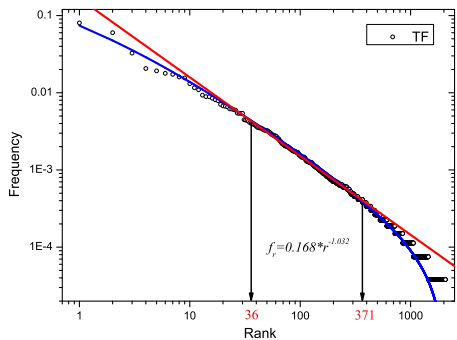}
\caption{ (Color online)
Frequency vs. rank for the text TF; see Table I. Red line: the Zipf curve $f_r=0.168 r^{-1.032}$.
Arrows indicate on the validity range of the Zipf's law.
Blue line: the solution of (\ref{g4}, \ref{g5}) for $c=0.168$ and $n=2067$. It coincides with the generalized Zipf 
law (\ref{g6}) for $r>r_{\rm min}=36$. The step-wise behavior of $f_r$ for $r>r_{\rm max}$ refers to hapax legomena.
} 
\vspace{0.25cm}
\label{f3}
\end{figure}

{\bf 1.} For each text there is a specific (Zipfian) range of ranks
$r\in [r_{\rm min}, r_{\rm max}]$, where the Zipf's law holds with
$\gamma\approx 1$ and $c< 0.2$ \cite{zipfwiki,baa}; see Table I  and Fig.~1. 

{\bf 2.} Even if the same word enters into different texts it typically
has quite different frequencies there \cite{arapov}, e.g. among 83
common words in the Zipfian ranges of AR and DL [see Table~I], only 12
words have approximately equal ranks and frequencies. 

{\bf 3.} The pre-Zipfian $1\leq r<r_{\rm min}$ range contains mainly
function words. They serve for establishing grammatical constructions
(e.g., {\it the, a, such, this, that, where, were}). But the majority of
words in the Zipfian range do have a narrow meaning (content words). A
subset of those content words has a meaning that is specific for the
text and can serve as its keywords \cite{ibm}. Below [in {\bf 15}]
we explain why the key-words appear in the Zipfian
domain. 

{\bf 4.} The absolute majority of different words with ranks in $[r_{\rm
min}, r_{\rm max}]$ have different frequencies.  Only for $r\simeq
r_{\rm max}$ the number of different words having the same frequency is
$\simeq 10$. For $r> r_{\rm max}$ we meet the {\it hapax legomena}:
words occuring only few times in the text ($f_rN=1,2,...$ is a small integer),
and many words having the same frequency $f_r$ \cite{baa}. The
effect is not described by a smooth rank-frequency relation, including
the Zipf's law. 

{\bf 5.} The minimal frequency of the Zipfian domain holds
$f_{r_{\rm max}} > {c}/{n}$.
We checked that this is valid not only for separate texts but also
for the frequency dictionaries of English and Irish. 
For our texts a stronger relation holds $f_{r_{\rm max}} \gtrsim \frac{1}{n}$. 
Hence $f_{r_{\rm max}}N \gtrsim \frac{N}{n}\gg 1$; see Table I. 
 
\textit{\textbf{Introduction to the model}}. A model for
the Zipf's law is supposed to satisfy the following features.  

({\bf I}) Apply to separate texts, i.e. explain how different texts can
satisfy the same form of the rank-frequency relation despite the fact
that the same words do {\it not} occur with same frequencies in the
different texts; see {\bf 2}.

({\bf II}) Derive the law together with its extensions for all frequencies, limits of validity and hapax legomena effect.

({\bf III}) Relate the law to formation of a text. 

Two sources of the model are the latent semantic analysis \cite{hof}, and the idea of applying ordered
statistics for rank-frequency relations \cite{dunaev,gusein,pietro}. 

Our model makes four (${\bf A}-{\bf D}$) assumptions. 

{\bf A.} The {\it bag-of-words picture} focusses on the frequency of the words that occur in a text
and neglects their mutual disposition (i.e. syntactic structure) \cite{madsen}.
Given $n$ different words $\{w_k\}_{k=1}^n$, the joint probability for $w_k$ to occur $\nu_k\geq 0$
times in a text $T$ is multinomial
\BEA
\label{multinom}
\pi[\nunu |\ff]=\frac{N!\,\f_1^{\nu_1}...\f_n^{\nu_n}}{\nu_1!...\nu_n!}, ~~~\nunu = \{\nu_k\}_{k=1}^{n}, ~~ \ff=\{\f_k\}_{k=1}^{n},
\EEA
where $N=\sum_{k=1}^n\nu_k$ is the length of the text, $\nu_k$ is the number of occurrences of $w_k$, 
and $\f_k$ is the
probability of $w_k$. The picture is well-known in computational 
linguistics \cite{madsen}. But for our purposes it incomplete, because
it implies that each word has the same probability 
for different texts [recall ({\bf I})]. 

{\bf B.} To improve this point we make $\ff$ a random vector
\cite{madsen} with a text-dependent density $P(\ff|T)$. The simplest
assumption is that $(T,\ff,\nunu)$ form a Markov chain: the 
text $T$ influences the observed $\nunu$ only via $\ff$. Then the
probability $p(\nunu|T)$ of $\nunu$ in a given text $T$ reads
\BEA
p(\nunu|T)={\int}\d \ff \, \pi[\nunu |\ff]\, P(\ff |T).
\label{bay}
\EEA
This form of $p(\nunu|T)$ is basic for probabilistic latent semantic
analysis \cite{hof}, a successful method of computational linguistics.
There the density $P(\ff |T)$ of latent variables $\ff$ is determined
from the data fitting. But we shall deduce $P(\ff |T)$ theoretically. 

{\bf C.} $P(\ff |T)$ is generated from a density $P(\ff)$
via conditioning on the ordering of $\w=\{w_k\}_{k=1}^n$ in $T$:
\BEA
\label{chua}
P(\ff |T)= P(\ff)\,\chi_T(\ff,\w)\left/ {\int}\d \ff' \, P(\ff')\, \chi_T(\ff',\w)\right. .
\EEA
If different words of $T$ are ordered as $(w_1,...,w_n)$ with respect to
the decreasing fequency of their occurence in $T$ (i.e. $w_1$ is more
frequent than $w_2$), then $\chi_T(\ff,\w)=1$ if $\f_1\geq...\geq\f_n$,
and $\chi_T(\ff,\w)=0$ otherwise. 

As substantiated below, $P(\ff)$ refers to the mental lexicon of the author prior to generating
a concrete text. 

{\bf D.} For simplicity, we assume that the probabilities $\f_k$ are
distributed identically and the dependence among them is due to
$\sum_{k=1}^n\f_k=1$ only:
\BEA
\label{g1}
P(\ff)\propto u(\f_1)\,...\,u(\f_n)\,\delta({\sum}_{k=1}^n\f_k-1),
\EEA
where $\delta(x)$ is the delta function and the normalization 
ensuring $\int_0^\infty\prod_{k=1}^n \d \f_k\, P(\ff)=1$ is omitted.

\textit{\textbf{Solution of the model and the Zipf's law}}.  The conditional
probability $p_r(\nu|T)$ for the $r$'th most frequent word $w_r$ to occur $\nu$ times in the 
text $T$ reads from (\ref{multinom}, \ref{bay})
\BEA
\label{g02}
p_r(\nu|T)&=&\frac{N!}{\nu!(N-\nu)!}\int_0^1\d \f\, \f^{\nu}(1-\f)^{N-\nu}P_r(\f|T),\\
\label{g2}
P_r(t|T)&=&\int \d\ff\, P(\ff|T)\delta(t-\f_{r}),
\EEA
where $P_r(t|T)$ is the marginal density for the probability $t$ of $w_r$.
For $n\gg 1$, we deduce from (\ref{chua}, \ref{g1}) that $P_r(t|T)$
follows the law of large numbers \cite{sup}. It is Gaussian,
\BEA
\label{gaugau}
P_r(t|T)\propto \exp[-\frac{n^3}{2\sigma_r^2}(t- \phi_r )^2],
\EEA
where $\sigma_r={\cal O}(1)$ [for $\phi_r={\it o}(1)$], and the mean $\phi_r$ is found from
two equations for two unknowns $\zmu$ and $\phi_r$:
\BEA
\label{g4}
{r}/{n}=
\int_{\phi_r}^\infty \d \f\,u(\f)\, e^{-\zmu \f n}\left /\int_{0}^\infty \d \f\, u(\f)\,e^{-\zmu \f n}\right. , \\
\label{g5}
\int_{0}^\infty \d \f\,\f\, u(\f)\, e^{-\zmu \f n}=\frac{1}{n}\int_{0}^\infty \d \f\,u(\f)\, e^{-\zmu \f n}.
\EEA
Eq.~(\ref{gaugau}) holds for $P_r(t|T)$ whenever its standard deviation 
$\sigma_r n^{-3/2}$ is much smaller than the mean $\phi_r$; as checked below,
this happens already for $r>10$.

{\bf 6.}
The meaning of (\ref{g4}, \ref{g5}) is explained via the marginal
density $P(\f_1)=\int_0^\infty\prod_{k=2}^n\d \f_k \, P(\ff)\propto
u(\f_l)e^{-\zmu \f_ln}$ found from (\ref{g1}) \cite{sup}.
Eq.~(\ref{g5}) ensures that ${\int}_0^\infty \d\f \,
\f\,P(\f)=\frac{1}{n}$. This relation follows from $\sum_{k=1}^n\f_k=1$
and it determines $\zmu$, an analogue of the chemical potential in
statistical physics \cite{sup}. The interpretation of (\ref{g4}) is that
it equates the relative rank $r/n$ to the (unconditional) probability
${\int}_{\phi_r}^\infty \d\f \, P(\f)$ of $\f\geq \phi_r$. 

Let us study implications of (\ref{g02}--\ref{g5}) for the Zipf's law. 

{\bf 7.} In (\ref{g02}), $P_r(\f|T)$ is much more narrow peaked than
$\f^{\nu}(1-\f)^{N-\nu}$, since $n^3\gg N\gg 1$ [see Table~I]. Hence in
this limit we approximate $P_r(\f|T)$ by delta-function
$\delta(\f-\phi_r)$ [see (\ref{gaugau})]:
\BEA
\label{g00}
p_r(\nu|T)=\frac{N!}{\nu!(N-\nu)!} \phi_r^{\nu}(1-\phi_r)^{N-\nu}.
\EEA
Eq.~(\ref{g00}) is the main outcome of the model; it shows that the conditional
probability $p_r(\nu|T)$ for the occurence number $\nu$ of the word
$w_r$ has the same form (\ref{g00}) for different text (see {\bf I}). In (\ref{g00}), $\phi_r$
is the effective probability of the word $w_r$. If 
$N\phi_r\gg 1$, $p_r(\nu|T)$ is peaked at $\nu=N\phi_r$: the frequency of a word that appears many times
equals its probability.  Each word of the Zipfian domain occurs at least
$\nu\sim N/n\gg 1$ times; see {\bf 5}. For such words we approximate
$f_r\equiv\nu/N\simeq\phi_r$.

{\bf 8.} Now we postulate in (\ref{g1})
\BEA 
\label{g3} 
u(f)=({n}^{-1}{c}+f)^{-2},
\EEA
where $c$ is related below to the prefactor of the Zipf's law.
Eq.~(\ref{g3}) is explained in {\bf 13-15} below.

{\bf 9.}
For $c\lesssim 0.2$, $c\zmu$ determined from (\ref{g5}, \ref{g3}) is small 
and is found from integration by parts: 
\BEA
\label{simmens}
\zmu\simeq c^{-1}\,e^{-\gamma_{\rm E}-\frac{1+c}{c}},
\EEA
where $\gamma_{\rm E}=0.55117$ is the Euler's constant.  One solves
(\ref{g4}) for $c\zmu\to 0$: $\frac{r}{n}=ce^{-n \phi_r\zmu}/(c+n
\phi_r)$.  For $r>r_{\rm min}$, $\phi_rn\zmu=f_rn\zmu<0.04\ll 1 $; see
(\ref{simmens}) and Table~I. We get
\BEA
\label{g6}
f_r=c(r^{-1}-n^{-1}).
\EEA
This is the Zipf's law generalized by the factor $n^{-1}$ at high ranks
$r$. This cut-off factor ensures faster [than $r^{-1}$] decay of $f_r$
for large $r$.  In literature a cut-off factor similar to $\frac{1}{n}$
is introduced due to additional mechanisms (hence new
parameters); see \cite{zane}. In our situation the power-law and cut-off
come from the same mechanism. 

Fig.~1 shows that (\ref{g6}) reproduces well the empirical behavior of
$f_r$ for $r>r_{\rm min}$. Our derivation shows that $c$ is the
prefactor of the Zipf's law, and that our assumption on 
$c< 0.2$ above (\ref{simmens}) agrees with observations; see Table~I.
For $c\gg 0.2$, (\ref{g4}, \ref{g5}) do not predict the
Zipf's law (\ref{g6}). 

{\bf 10.} For given prefactor $c$ and the number of different words $n$,
(\ref{g4}--\ref{g3}) predict the Zipfian range $[r_{\rm min},r_{\rm
max}]$ in agreement with empirical results; see Fig.\,1. 

{\bf 11.} For $r<r_{\rm min}$, it is not anymore true that $f_rn\zmu\ll 1$. So
the fuller expression (\ref{g4}) is to be used. It reproduces qualitatively
the empiric behavior of $f_r$; see Fig.~1. 


{\bf 12.} According to (\ref{g00}), the probability $\phi_r$ is small for $r\gg r_{\rm
max}$ and hence the occurence number $\nu\equiv f_rN$ of a words $w_r$ is a
small integer (e.g. 1 or 2) that cannot be approximated by a continuous
function of $r$; see (\ref{g3}) and Fig.~1.  To describe this hapax
legomena range, define $r_k$ as the rank, when $\nu\equiv f_rN$ jumps from
integer $k$ to $k+1$. Since $\phi_r$ reproduces well the trend of $f_r$
even for $r>r_{\rm max}$, see Fig.~1, $r_k$ can be theoretically predicted from (\ref{g6})
by equating its left-hand-side to $k/N$:
\BEA
\hat{r}_k=[\frac{k}{Nc}+\frac{1}{n}]^{-1}, \qquad k=0,1,2,...
\label{hapo}
\EEA
Eq.~(\ref{hapo}) is exact for $k=0$, and agrees with $r_k$ for $k\geq
1$; see Table II. Hence it describes the hapax legomena phenomenon (many
words have the same small frequency) \cite{f_hapax}.

\begin{table}
\begin{center}
\tabcolsep0.03in \arrayrulewidth0.5pt
\renewcommand{\arraystretch}{1.6}
\caption{Description of the hapax legomena for the text TF; see Table~I and (\ref{hapo}).
The maximal relative error $\frac{\hat{r}_k-r_k}{r_k}=0.0357$ is reached for $k=6$.}
\begin{tabular}{|c|c|c|c|c|c|c|c|c|c|c|}
\hline
{$r$}/{$k$} & 1 & 2& 3 & 4&5&6 & 7&8&9&10\\
\hline
 $r_k$ & 1446 & 1061 &848 & 722 & 611 & 529& 474& 437&398&370 \\
\hline
$\hat{r}_k$ & 1414 & 1074 & 866 & 726 & 624&547& 488&440&400&368 \\
\hline
\end{tabular}
\end{center}
\end{table}

\textit{\textbf{Preliminary summary}}. Thus {\bf 9-12} achieved the
promises ({\bf I}) and ({\bf II}) of our program: though different texts
can have different frequencies for same words, the frequencies of words
in a given text follow the Zipf's law with the correct prefactor
$c\lesssim 0.2$. Without additional fitting parameters and new
mechanisms we recovered the corrected form of this law applicable for
large and small frequencies [see {\bf 11}, {\bf 12}]. But why we would
select (\ref{g3}), if we would not know that it reproduces the Zipf's
law? Answering this question will fulfil ({\bf III}). 

\textit{\textbf{Mental lexicon and the apriori density}}. 
Here we explain the choice (\ref{g1}, \ref{g3}) for the apriori
probability density for the probabilities $\ff=(\f_1,...,\f_n)$ of
different words $(w_1,...,w_n)$.  To avoid the awkward term ``probability
for probability" we shall call $P(\ff)$ likelihood. We focus on the
marginal likelihood [see {\bf 6} and (\ref{g3})]:
\BEA
\label{pri}
P(\f)=(n^{-1}{c}+\f)^{-2}e^{-\mu n \f},
\EEA
since $P(\f)$ determines the rank-frequency
relation (\ref{g4}). For a more detailed discussion 
of the items below see \cite{sup}.

{\bf 13.} The basic reason for the words to have random (variable)
probabilities is that the text-producing author should be able to
compose different texts, where the same word can have very different
frequencies [see {\bf I}]. Hence $P(\ff)$ relates to the prior knowledge
(or lexicon) of the author on words. This concept of {\it mental
lexicon} is an established one in psycholinguistics \cite{mental,levelt}. 

{\bf 14.} Once each word $w_k$ has to have a variable probability
$\f_k$, there should be a way for the author to increase it, e.g. when the
authors decides that $w_k$ should become a keyword of the text. The
ensuing relation between the probability vectors $\ff'$ (new) and $\ff$
(old) should be a group, since the author should be able to come back
from $\ff'$ to $\ff$ when revising the text. Under certain natural
conditions, the only such group with parameters $\tau_k$
is \cite{jaeger}:
\BEA
\label{s2}
\f_k'= {\tau_k \f_k} \,\left[{{\sum}_{l=1}^n \tau_l \f_l}\right]^{-1} , ~~ \tau_k>0, ~~ k=1,...,n,
\EEA
Eq.~(\ref{s2}) is a
generalized Bayes formula \cite{jaeger,sup}. It is used in the Bayesian statistics for
motivating the choice of priors \cite{jaeger}, a task related to ours.

If the author wants to increase 
$\tau_1$ times the probability of the word $w_1$, then in (\ref{s2})
$\tau_1>1$ and $\tau_{k\geq 2}=1$:
\BEA
\label{gro}
\f_1'= \frac{\tau_1 \f_1}{1+(\tau_1-1)\f_1}, ~ \f_l'= \frac{ \f_l}{1+(\tau_1-1)\f_1}, 
~ {\rm for}~ l\geq 2. 
\EEA
The inverse of (\ref{gro}) is found by interchanging $\f_k'$ with $\f_k$ and
$\tau_1$ with $\tau_1^{-1}$. For the Zipf's law the relevant probabilities are small,
$\theta_1'< {\cal O}(1/n)$; see {\bf 9} and Fig.~1. Then
$1+(\tau_1^{-1}-1)\f_1'\approx 1$ and (\ref{gro}) becomes the
scaling transformation of one variable: $\f_1'=\tau_1 \f_1$,
$\f_l'=\f_l$, $l\geq 2$. The new likelihood reads from (\ref{gro},
\ref{pri})
\BEA
\label{sos}
P'(\f_1')=\frac{1}{\tau_1}P(\frac{\f_1'}{\tau_1})=\frac{1}{\tau_1}(\frac{c}{n}+\frac{\f_1'}{\tau_1})^{-2}.
\EEA
Other densities do not change $P'(\f_l')=P(\f_l')$ for $l\geq 2$.

{\bf 15.} Once $P(\f)$ describes the mental lexicon, and (\ref{s2}) is
an operation by which the text is written, we suppose that the features
of $P(\f)$ can be explained by checking its response to (\ref{s2}).  For
the ratio of the new to the old likelihood of the probability $\f_1'$ we
get from (\ref{sos})
\BEA
\label{kopt1}
P'(\f_1')/P(\f_1')&=&\tau_1>1 ~~ {\rm for}~~ \f_1'\gg c\tau_1/n , \\
                  &=&\tau_1^{-1}<1 ~~ {\rm for}~~ \f_1'\ll c\tau_1/n.
\label{kopt2}
\EEA
The meaning of (\ref{kopt1}, \ref{kopt2}) is that once the author decides to increase
the probability of the word $w_1$ by $\tau_1$ times, this word will be
$\tau_1$ times more likely produced with the higher probabilities, and
$\tau_1$ times less likely with smaller probabilities; see
(\ref{kopt2}). This is the mechanism that ensures the appearance of the keywords 
in the Zipfian range. It is unique to the form (\ref{pri}) of the
marginal likelohood, which by itself is due to the form (\ref{g3}) of
$u(\f)$. 

If $P(\f)$ is assumed to reflect the organization of the mental lexicon,
then according to (\ref{kopt1}, \ref{kopt2}) this organization is
efficient, because the decision on increasing the probability of $w_1$
translates to increasing the likelihood of larger values of the
probability. The organization is also stable, since the likelihood at
large probabilities increases right at the amount the author planned,
not more. 

\textit{\textbf{Conclusion}}. We answer the first question asked in the
introduction: the Zipf's law|together with the limits of its validity,
its generalization to high and low frequencies and hapax
legomena|relates to the stable and efficient organization of the mental
lexicon of the text-producing author. Practically, our derivation of the
Zipf's law will motivate the usage of prior (\ref{g3}) in the schemes of
latent semantic analysis. We expect these schemes to be more efficient
for real texts, if the prior structure of the model conforms the Zipf's
law. The proposed methods can find applications for studying
rank-frequency relations and power laws in other fields. 

We thank A. Galstyan and D. Manin for discussions. This work is supported by the Region des Pays 
de la Loire under the Grant 2010-11967.

\pagebreak

\begin{center}
{\large \bf Supplementary Material}
\end{center}

\vspace{1cm}

In this supplementary material to the main text we review the linear
fitting method, derive and clarify Eqs.(8-10) from the section
\textit{\textbf{Solution of the model and the Zipf's law}} of the main
text, derive the expression for the marginal probability [Eq.~(16) and
point {\bf 6} of the main text], and discuss in more detail the content
of section \textit{\textbf{Mental lexicon and the apriori density}}.
These tasks are carried out in, respectively, sections I, II, III and IV
below. 

\section{I. Linear fitting}
\label{linear}

Here we recall the main ideas of the linear fitting method that is
employed in the section \textit{\textbf{The validity range of the Zipf's
law}} of the main text. 

Table I of the main text presents 3 texts we studied (we worked out more
texts that consistently show the same applicability pattern of the
Zipf's law). For each text we extract the ordered frequencies of
different words [the number of different words is $n$; the overall
number of words in a text is $N$]:
\BEA 
\label{001} \{f_r\}_{r=1}^{n}, ~~ f_1\geq ...\geq f_{n}, ~~
{\sum}_{r=1}^{n} f_r =1.  
\EEA
We should now see whether the data $\{f_r\}_{r=1}^{n}$ fits to a power law: $\hat{f}_r=cr^{-\gamma}$.
We represent the data as
\BEA
\{y_r(x_r)\}_{r=1}^{n}, ~~ y_r=\ln f_r, ~~ x_r=\ln r, 
\EEA
and fit it to the linear form $\{\hat{y}_r=\ln c-\gamma x_r\}_{r=1}^{n}$. Two 
unknowns $\ln c$ and $\gamma$ are obtained from minimizing the sum of
squared errors:
\BEA
SS_{\rm err}={\sum}_{r=1}^{n} (y_r-\hat{y}_r)^2.  
\EEA
It is known since Gauss that this minimization produces
\BEA
\label{laude}
-\gamma^* = \frac{ \sum_{k=1}^n (x_k-\overline{x})(y_k-\overline{y})    }{ \sum_{k=1}^n (x_k-\overline{x})^2}, ~~
\ln c^*=\overline{y}+\gamma^*\overline{x},
\EEA
where we defined
\BEA
\overline{y}\equiv \frac{1}{n}{\sum}_{k=1}^n y_k, ~~
\overline{x}\equiv \frac{1}{n}{\sum}_{k=1}^n x_k.
\EEA
As a measure of fitting quality one can take:
\BEA
{\rm min}_{c,\gamma}[SS_{\rm err}(c,\gamma)]=SS_{\rm err}(c^*,\gamma^*)\equiv SS_{\rm err}^*.
\EEA
This is however not the only relevant quality measure. Another (more global) aspect of this quality is
the coefficient of correlation between $\{y_r\}_{r=1}^{n}$ and $\{\hat{y}_r\}_{r=1}^{n}$ \cite{lili1}:
\BEA
R^2=\frac{  \left[\, \sum_{k=1}^n (y_k-\bar{y}) (\hat{y}^*_k-\overline{\hat{y}^*})\, \right]^2    }{\sum_{k=1}^n (y_k-\bar{y})^2
\sum_{k=1}^n (\hat{y}^*_k-\overline{\hat{y}^*})^2},
\EEA
where 
\BEA
\hat{y}^*=\{\hat{y}^*_r=\ln c^*-\gamma^* x_r\}_{r=1}^{n}, ~~~
\overline{\hat{y}^*}\equiv \frac{1}{n}{\sum}_{k=1}^n \hat{y}^*_k.
\EEA
For the linear fitting (\ref{laude}) the squared correlation coefficient is equal to the coefficient of
determination, 
\BEA
R^2= {\sum}_{k=1}^n (\hat{y}^*_k-\overline{y})^2\left/ {\sum}_{k=1}^n(y_k-\overline{y})^2, \right.
\EEA
the amount of variation in the data explained by the fitting
\cite{lili1}. Hence $SS_{\rm err}^*\to 0$ and $R^2\to 1$ mean good
fitting. We minimize $SS_{\rm err}$ over $c$ and $\gamma$ for $r_{\rm
min}\leq r\leq r_{\rm max}$ and find the maximal value of $r_{\rm
max}-r_{\rm min}$ for which $SS_{\rm err}^*$ and $1-R^2$ are smaller
than, respectively, $0.05$ and $0.005$. This value of $r_{\rm
max}-r_{\rm min}$ also determines the final fitted values $c^*$ and
$\gamma^*$ of $c$ and $\gamma$, respectively; see Tables I, II and
Fig.~1. Thus $c^*$ and $\gamma^*$ are found simultaneously with the
validity range $[r_{\rm max},r_{\rm max}]$ of the law.  Whenever there
is no risk of confusion, we for simplicity refer to $c^*$ and $\gamma^*$
as $c$ and $\gamma$, respectively.

\section{II. Derivation of Eqs.~(8-10) of the main text.}

In (7) of the main text we defined $P_r(t|T)$: the marginal density for 
the probability $t$ of the word $w_r$. Using (4,5) of the main text, we rewrite 
(7) of the main text as
\BEA
P_r(t|T)
&&\propto\int_0^\infty \d \f_1\int_0^{\f_1}\d \f_2\int_0^{\f_2}\d \f_3...\int_0^{\f_{n-1}}\d \f_{n}\times\nonumber\\
&&P(\f_1,...,\f_n)\,\delta(t-\f_{r}),
\label{ggg2}
\EEA
where 
\BEA
\label{ggg1}
P(\ff)\propto u(\f_1)\,...\,u(\f_n)\,\delta({\sum}_{k=1}^n\f_k-1), 
\EEA
as given by (7) of the main text. Recall that $\ff=(\f_1,...,\f_n)$.

In (\ref{ggg1}) we employ the Fourier representation of the delta-function,
\BEA
\label{star}
\delta({\sum}_{k=1}^n\f_k-1)=\int_{-i\infty}^{i\infty} \frac{\d z}{2\pi i}\, e^{z-z{\sum}_{k=1}^n\f_k},
\EEA
put (\ref{ggg1}) into (\ref{ggg2}) and then apply integration by parts. The result reads
\begin{gather}
\label{a1}
P_r(t|T)
\propto u(t)
\int_{-i\infty}^{i\infty} \frac{\d z\, e^{z}}{2\pi i} \chi_0^{n-r}(t,z)\chi_1^{r-1}(t,z)\,e^{-tz},
\end{gather}
where
\begin{gather}
\chi_0(t,z)\equiv\int_0^t\d y e^{-zy}u(y),~
\chi_1(t,z)\equiv\int_t^\infty\d y e^{-zy}u(y).\nonumber
\end{gather}
The integral in (\ref{a1}) will be worked out via the saddle point method. 
But before that we need to fix the scales of the involved quantities. To this end,
make the following changes of variables
\BEA
\label{a3}
\bz =z/n, ~~ \bt =tn, ~~ \by =yn,~~ \br =r/n.
\EEA
Then $P_r(t|T)$ reads from (\ref{a1})
\BEA
\label{a4}
P_r(t|T) &\propto&  u(t)\int_{-i\infty}^{i\infty}\frac{\d \bz}{2\pi i}\,e^{n\varphi(\bt,\bz)-\bt\bz},\\
\varphi(\bt,\bz)&=& \bz+(1-\br)\ln \int_0^{\bt} \frac{\d y\, e^{-\bz y} } { (c+y)^2}\nonumber\\
&+& (\br-\frac{1}{n})\ln\int_{\bt}^\infty\frac{\d y \, e^{-\bz y} }{ (c+y)^2},
\label{urkan}
\EEA
where in (\ref{urkan}) we already used $u(t)=(n^{-1}c+t)^{-2}$; see (12) of the main text. 

If $n\gg 1$ and $0<\br<1$ is a finite number (neither close to one, nor to zero), 
the behavior of $\rho_r(t)$ in various averages, e.g. $\int
\d t\, t\, \rho_r(t) $, is determined by the values of $\bz=\bz_s$ and
$\bt=\bt_s$ that maximize $\phi(\bt,\bz)$. They are found from
saddle-point equations
\BEA
\label{gado}
\partial_{\bt} \phi(\bt_s,\bz_s)=\partial_{\bz} \phi(\bt_s,\bz_s)=0.
\EEA
After reworking the two equations (\ref{gado}) we get Eqs.~(9,10) of the main text. 

Due to (\ref{a3}), $\bz_s$ (that is real and positive) and $\bt_s$
stay finite for $n\gg 1$. Hence the integration line over $\bz$ in
(\ref{a4}) is shifted to pass through $\bz_s$ (the saddle-point method).
Now $\phi(\bt,\bz)$ is expanded around $\bz=\bz_s$ and $\bt=\bt_s$
[first-order terms nullify due to (\ref{gado})]:
\BEA
\label{rimma}
&&\phi(\bt,\bz)=\phi(\bt_s,\bz_s)+\frac{1}{2}\partial_{\bt\bt} \phi(\bt_s,\bz_s)(\bt-\bt_s)^2 \\
&&+\frac{1}{2}\partial_{\bz\bz} \phi(\bt_s,\bz_s)(\bz-\bz_s)^2\\
&&+\partial_{\bt\bz} \phi(\bt_s,\bz_s)(\bt-\bt_s)(\bz-\bz_s)+....
\EEA
Now only these terms can be retained in the integral over $\bz$. Since
this integral goes over the imaginary axis, while $\bz_s$ is real, the
integration contour is to be shifted to pass through $\bz_s$. For the
convergence of the resulting Gaussian integral we need
$\frac{1}{2}\partial_{\bz\bz} \phi(\bt_s,\bz_s) >0$. Taking this Gaussian integral leads us
to [up to factors that either constant or irrelevant for $n\gg 1$]
\BEA
\label{gaugau1}
P_r(t|T) \propto e^{-\frac{n}{2\sigma^2}(\bt-\bt_s)^2}=
e^{-\frac{n^3}{2\sigma^2}(t-\frac{\bt_s}{n})^2},\\
\label{mundar}
\frac{1}{\sigma^2} = \frac{[\,\partial_{\bt\bz} \phi(\bt_s,\bz_s)\,]^2}{\partial_{\bz\bz} \phi(\bt_s,\bz_s)}-\partial_{\bt\bt} \phi(\bt_s,\bz_s).
\EEA
Hence $P_r(t|T)$ is approximately Gaussian,
with the standard deviation ${\cal O}(n^{-3/2})$ much smaller
than the average for $\bt_s={\cal O}(1)$. 

In working out (\ref{mundar}), we shall
employ the fact that in (\ref{urkan}) $\bz_s=\mu$ is a
small parameter; see (13) of the main text. 
This produces [up to smaller corrections]
\BEA
\label{gr}
\sigma=(c+\bt_s)\sqrt{\bt_s}.
\EEA
Eq.~(\ref{gaugau1}) derives (8) of the main text, while (\ref{gr}) 
accounts for the estimate of $\sigma$ that was presented after
(8) of the main text.

\section{III. Derivation of the marginal probability (Eq.~(16) and point {\bf 6} of the main text).}

The marginal probability $P(t)$ is defined from (\ref{ggg1}) as
\BEA
\label{osh}
P(t)=\int \d\ff P(\ff)\,\delta(t-\f_{r}).
\EEA
using (\ref{ggg1}, \ref{star}) we obtain from (\ref{osh})
\BEA
\label{a5}
P(t) \propto  u(t)\int_{-i\infty}^{i\infty}\frac{\d \bz}{2\pi i}\,e^{n\phi(t,\bz)-\bt\bz},\\
\phi(t,\bz)=(1-t) \bz+\ln \int_0^{\infty} \d y\, e^{-\bz y}\, (c+y)^{-2}.
\EEA
We use the saddle-point method for (\ref{a5}). This produces the same
saddle-point equation (\ref{gado}) for $\bz_s$, 
\BEA
1=\frac{\int_0^{\infty} \d y\, e^{-\bz_s y}\, (c+y)^{-2}}{\int_0^{\infty} \d y\,y\, e^{-\bz_s y}\, (c+y)^{-2}},
\EEA
provided that we note the dominant range $t\propto 1/n\ll 1$ of $t$. Thus 
\BEA
P(\f) \propto u(\f)e^{-n\f\bz_s}. 
\label{glan}
\EEA
This validates Eq.~(16) of the main text, as well as its point {\bf 6}.

Likewise, one can show that the marginal density
$P(\f_1,...,\f_m)$ factorizes provided that $m\ll n$:
\BEA
\label{g111}
P(\f_1,...,\f_m)\propto u(\f_1)e^{-\zmu \f_1n}\,...\,u(\f_m)e^{-\zmu \f_m n}.
\EEA
Eq.~(\ref{g111}) can be established more heuristically via the exact relation $\overline{
\,[\sum_{k=1}^n \f_k]^2\,}=1$, where $\overline{\, f\,}$ means averaging
over $P(\f_1,...,\f_n)$.  This relation predicts, together with
$\overline{\, \f_k\,}=\frac{1}{n}$, that $\overline{\,
\f_i\f_j\,}-\overline{\, \f_i\,}\, \overline{\, \f_j}={\cal O}(n^{-3})$,
hence approximate factorization.

Using (\ref{glan})
with $u(\f)=(\frac{c}{n}+\f)^{-2}$ we note that the standard deviation
$\langle(\f-\langle \f\rangle)^2\rangle=\frac{1}{n}\sqrt{\frac{c}{\bz_s}-1}\simeq
\frac{1}{n}\sqrt{\frac{c}{\bz_s}}$ is larger than the average
$\langle \f\rangle=\int\d\f \f P(\f)=\frac{1}{n}$, since $c/\bz_s\gg 1$.

\section{IV. Mental lexicon and apriori density}

This is an expanded version of the coresponding section of the main text. 
We explain the choice
\begin{gather}
\label{ml1}
P(\ff)\propto u(\f_1)\,...\,u(\f_n)\,\delta({\sum}_{k=1}^n\f_k-1), \\ 
u(\f)=({c}{n}^{-1}+\f)^{-2},
\label{ml2}
\end{gather}
for the apriori probability density for the probabilities $\ff=(\f_1,...,\f_n)$
of different words $w_1,...,w_n$.
To avoid the awkward term ``probability for probability" we shall call $P(\ff)$
likelihood.

Recall that the marginal likelihood deduced from (\ref{ml1}) reads
\BEA
\label{ml4}
P(\f)=(n^{-1}{c}+\f)^{-2}e^{-\mu n \f},
\EEA
where $\mu$ is determined by (12,16) of the main text.

We shall explain the choice (\ref{ml2}) via the features of the marginal
likelihood (\ref{ml4}), because it eventually determines the
rank-frequency relation leading to the Zipf's law. 

The numbering of the items {\bf 13-15} below coincides that in the
section \textit{\textbf{Mental lexicon and the apriori density}} of the
main text.  The items {\bf 13.2}, {\bf 13.3}, {\bf 13.4}, {\bf 16} and
{\bf 17} below are added additionally, they are absent in the main text. 

{\bf 13.1} Recall that the basic reason for the words to have random
(not fixed) probabilities is that the text-producing author should be
able to compose different texts, where the same word can have different
frequencies. Hence the likelihood $P(\ff)$ of random probabilities
relates to the prior knowledge (or lexicon) of the text-generating
author on the words.  This concept of {\it mental lexicon}|the store of
words in the long-time memory so that the words are employed on-line for
expressing thoughts via phrases and sentences|is well-established in
psycholinguistics \cite{mental1}. Though there is no a unique theory of
mental lexicon|there is only a diverse set of competing models
\cite{mental1}|some of its basic features are well-established
experimentally and are employed below for explaining the choice
(\ref{ml1}, \ref{ml2}). 

{\bf 13.2} We assume that during the conceptual planning of the text,
i.e. when deciding on its topic, style and potential audience, the
author already chooses (at least approximately) two structural
parameters: the number $n$ of different words to appear there and the
constant $c$. This is why the marginal likelihood (\ref{ml4}) depends on
the parameters $c$ and $n$.  We recall that $c$ (along with $n$) is a
structural parameter of the text, because according to the point {\bf 5} of the
main text, $c/n$ separates the Zipfian (keywords dominated) range from
the hapax legomena range (rare words). 

{\bf 13.3} Note that different words have the same marginal likelihood
(\ref{ml4}). Put differently, the likelihood $P(\ff)$ is symmetric with
respect to interchanging the words $w_1,...,w_n$. This feature relates
to an experimental fact that words are stored in the mental lexicon in
the same way \cite{levelt1}. The difference between them|e.g. whether the
word is more familiar to the author, and/or used by him more
frequently|can be relevant during the (later) phonologization stage of
speech/text production \cite{levelt1}; in this context
see also the item {\bf 17} below.

Naturally, the above symmetry holds for the apriori likelihood.  The
posterior likelihood $P(\ff|T)$ (see (6) of the main text), the one that
is conditioned over the written text, does not and should not have such
a symmetry.

{\bf 13.4}
Note that the marginal likelihood (\ref{ml4}) concentrates at small
probabilities $\f\simeq c/n$. The concentration holds locally|since
$P(\f)$ is peaked at $\f=0$ and is approximately constant for $\f\ll
c/n$| and {\it also} globally, i.e. on the level of the full probability:
\begin{gather}
\label{ml9}
{\rm Pr}[\f<a]=\int_{0}^{a}\d\f P(\f)=\int_a^1 \d\f P(\f)={\rm Pr}[\f>a], \\
{\rm for}~~ a\simeq c/n. \nonumber
\end{gather}
If $a$ is sufficiently larger (smaller) than $c/n$, the left-hand-side
of (\ref{ml9}) is larger (smaller) than its right-hand-side. 

The local and global concentrations are different from each other. For
example, consider $P(\f)\propto \f^{-1/2} e^{-\mu\f n}$. It displays a
local concentration around $\f\simeq 0$, but (\ref{ml9}) (global
concentration) predicts $a\simeq \frac{1}{\nu n}$. 

Hence according to (\ref{ml4}), apriori (i.e. before the text is
written) all the (content) words have small probabilities.  This is
explained as follows.  Since the majority of words in the mental lexicon
are potential keywords of some texts, apriori (i.e.  before the text is
written) they have small probabilities. Indeed, the defining (and
operationally used) feature of a keyword is that its frequency in a
given text is much larger than its frequency in a large mixture of
different texts \cite{keywords}. Thus the apriori likelihood of the
probability should be concentrated at small probabilities $\f\simeq
c/n$.

{\bf 14.} Once each word $w_k$ has to have a variable (random)
probability $\f_k$, there should be a way for the author to change
(increase or decrease) this probability, e.g. when the author decides
that the word $w_k$ is to become the keyword of the text. The ensuing
relation between the probability vectors $\ff'$ (new) and $\ff$ (old)
should be a group, since the author should be able to come back from
$\ff'$ to $\ff$, e.g.  when revising the text. 

One can impose two natural restrictions on this group \cite{jaeger1}.
These restrictions follow the general idea that the meaning of $\ff$ as
probabilities of certain events is conserved during the transformation. 

First, the words that have strictly zero probability $\f_k=0$ will stay
zero, i.e. $\f_k'=0$ if and only if $\f_k=0$. 

Second, the probability mixtures are conserved: if
\BEA
\ff =\lambda  \chichi +(1-\lambda) \etaeta, ~~ 0<\lambda<1,
\EEA
where $\chichi=(\chi_1,...,\chi_n)$ and $\etaeta=(\eta_1,...,\eta_n)$ are arbitrary probability vectors, and where
$\lambda$ is a (mixing) parameter, then 
\BEA
\ff' =\lambda \chichi'+(1-\lambda)  \etaeta'.
\EEA
Here primed and non primed probability vectors relate to each other via the sought group. 

The only group that (for $n\geq 3$) is consistent with the above two conditions is \cite{jaeger1}:
\BEA
\label{ml3}
\f_k'= \frac{\tau_k \f_k}{\sum_{l=1}^n \tau_l \f_l} , ~~ \tau_k>0, ~~ k=1,...,n,
\EEA
where $\tau_k$ are the group parameters. If the author wants to increase two times the probability of the word $w_1$, then 
$\tau_1=2$ and $\tau_{k\geq 2}=1$. 

Note that (\ref{ml3}) becomes the Bayes formula if we relate $\tau_k$ to
a conditional probability \cite{jaynes}. In this alternative
interpretation of (\ref{ml3}), the author has to
retrieve a word $w$ having certain specific features (i.e. it is a
transitive verb) from the set of words $w_1,...,w_n$ having
probabilities $\f_1,...,\f_n$. If we denote by ${\rm Pr}(E|w=w_k)$ the
conditional probability that the word $w_k$ displays the needed feature $E$,
we can relate in (\ref{ml3}) $\tau_k={\rm Pr}(E|w=w_k)$, and (\ref{ml3})
will describe the searching process for the word having the needed
feature $E$. 

{\bf 15.} Since $P(\ff)$ is the basic description of the mental lexicon
that enters into our model, and once (\ref{ml3}) is an operation by
which the text is ultimately written, it is natural to suppose that the
features of $P(\ff)$ can be explained by checking its response to
(\ref{ml3}).  It is with a similar purpose of motivating the prior
likelihood that (\ref{ml3}) is applied in Bayesian statistics
\cite{jaeger1,jaynes}. There, however, the attention is focused on the
non-informative prior likelihood that will stay invariant under 
(\ref{ml3}). This is not suitable for our purpose precisely because we
expect that the mental lexicon|whose organization $P(\ff)$ refers
to|will somehow reflect the basic mechanism (\ref{ml3}), i.e. $P(\f)$
will display specific changes under (\ref{ml3}).

In interpreting those changes, we adapt (\ref{ml3}) to the probability
increase of a single word $w_1$, whose probability the author decides
to increase by $\tau_1>1$ times.  Thus, (\ref{ml3}) is applied for
$\tau_2=...\tau_n=1$:
\BEA
\label{ml5}
\f_1'= \frac{\tau_1 \f_1}{1+(\tau_1-1)\f_1}, ~ \f_l'= \frac{ \f_l}{1+(\tau_1-1)\f_1}, 
~ {\rm for}~ l\geq 2. 
\EEA
The inverse of transformation (\ref{ml5}) reads
\BEA
\label{ml6}
\f_1= \frac{\tau_1^{-1} \f'_1}{1+(\tau_1^{-1}-1)\f'_1}, ~ \f_l= \frac{ \f'_l}{1+(\tau_1^{-1}-1)\f'_1}.
\EEA
In the frequency range we are interested in, $(\tau_1^{-1}-1)\f'_1$ can be neglected, hence (\ref{ml6})
just reduces to the scaling transformation:
\BEA
\label{ml7}
\f_1= \tau_1^{-1} \f'_1, ~ \f_l=  \f'_l.
\EEA
The change of the marginal likelihood for $\f_1$
is deduced from (\ref{ml4}, \ref{ml7}):
\BEA
\label{ml01}
P'(\f_1')=\frac{1}{\tau_1}P(\frac{\f_1'}{\tau_1})=\frac{1}{\tau_1}(\frac{c}{n}+\frac{\f_1'}{\tau_1})^{-2}.
\EEA
Thus, for the ratio of the new to the old likelihood of the probability $\f_1'$ we get
\BEA
\label{kopt11}
P'(\f_1')/P(\f_1')&=&\tau_1>1 ~~ {\rm for}~~ \f_1'\gg c\tau_1/n , \\
                  &=&\tau_1^{-1}<1 ~~ {\rm for}~~ \f_1'\ll c\tau_1/n.
\label{kopt21}
\EEA
The meaning of (\ref{kopt11}) is that once the author decides to increase
the probability of the word $w_1$ by $\tau_1$ times, this word will be
$\tau_1$ times more likely produced with the higher probabilities, and
$\tau_1$ times less likely with smaller probabilities; see
(\ref{kopt21}). The feature is unique to the form (\ref{ml4}) of the
marginal likelohood, which by itself is due to the form (\ref{ml2}) of
$u(\f)$. This is the mechanism that ensures the appearance of the keywords 
in the Zipfian range. 

If $P(\f)$ is assumed to reflect the organization of the mental lexicon,
then according to (\ref{kopt11}, \ref{kopt21}) this organization is
efficient, because the decision on increasing the probability of $w_1$
translates to increasing the likelihood of larger values of the
probability. The organization is also stable, because the likelihood at
large probabilities does increase right at that amount the author
planned (not more). 

{\bf 16.} Above we related the prior likelihood $P(\f)$ to the
organization of the mental lexicon. Now we would like to clarify this
relation by looking at some alternative forms of the marginal
likelihood, e.g. 
\BEA
\tilde{u}(\f)\propto (\tilde{c}n^{-1}+\f)^{-1},
\EEA
which will produce
\BEA
\label{obo}
\tilde{P}(\f)=(\tilde{c}n^{-1}+\f)^{-1} e^{-n\f\tilde{\mu} }.
\EEA
Here $\tilde{\mu}$ is determined from
\BEA
\int_{0}^\infty \frac{\d y (y-1)}{\tilde{c}+y}\, e^{-\tilde{\mu} y}=0,
\EEA
by analogy to (12) of the main text. 

It is clear that instead of (\ref{kopt11}), 
we now get $P'(\f_1')/P(\f_1')=1$, i.e the likelihood of large probabilities does not change at all. This indicates
on the lack of organization in the mental lexicon (or at least very inefficient organization).

The rank-frequency relation generated by (\ref{obo})
will read by analogy to (11) of the main text
\BEA
\label{g44}
\frac{r}{n}=\frac{\int_{\phi_rn}^\infty \frac{\d y}{\tilde{c}+y}\,
e^{-\tilde{\mu} y} }{\int_{0}^\infty \frac{\d y}{\tilde{c}+y}\,
e^{-\tilde{\mu} y}}.  
\EEA
In the limit of a sufficiently small $\tilde{c}$,
the rank-frequency relation obtained from (\ref{g44}) is exponential,
\BEA
\label{gorgi}
\phi_r\simeq {\alpha n}\,{e^{-\alpha n r}}, ~~\alpha = \ln(1/\tilde{c}),
\EEA
instead of the Zipf's law. According to (\ref{gorgi}) the majority of words have neglegible
frequencies, hence a small group of high-frequency words dominates the text. 
Intuitively, this connects well with the above statement on the lack of organization. 

{\bf 17.} The message of (\ref{kopt11}, \ref{kopt21}) closely relates (but
is not completely identical) to the word-frequency effect well-known for
the mental lexicon: more frequently used words are produced (recalled)
more easily \cite{mental1,levelt1,leod}. In the context of (\ref{kopt11},
\ref{kopt21}) this implies that the words that are decided to appear with
more probability (e.g. the keywords) will be more likely produced
with higher probabilities. 

Note that there is no contradiction between the message of (\ref{kopt11},
\ref{kopt21}) and the fact that all the words have the same marginal
apriori likelihood [see (\ref{ml4})]. The latter aspect refers to the word
as emerging from the mental lexicon, while the former implicitly refers
to the initial stages of writing the text. 

The same distinction is well known for the proper word-frequency effect
in speech production, i.e. producing words from the mental lexicon
\cite{levelt1}.  According to the accepted model \cite{levelt1} of this
process, during the first stage of speech production the author
conceptualizes his thought into the abstract form of the word (lemma).
This form reflects the meaning of the word and its syntactic usage, but
is not yet to be put in syllabic form and pronounced \cite{levelt1}.  The
word-frequency effect comes into play during this second stage, but is
absent when the lemma is activated in the mental lexicon \cite{levelt1}.
This is why the word-frequency can be even reversed|more frequent words
are recognized more easily|for those tasks (e.g. recognition) that
include mainly the lemma activation \cite{leod}.


\begin{thebibliography}{99}
\bibitem{f_hapax} For $k\gg Nc/n$ we deduce from (\ref{hapo})
$\hat{r}_k-\hat{r}_{k+1}\propto k^{-2}$ for the number of words having
the frequency $k/N$.  This relation, which is a crude particular case of
(\ref{hapo}), is sometimes called the second Zipf's law \cite{baa}.
\end{thebibliography}
\end{document}